\documentclass[11pt]{article}
\usepackage[cp1251]{inputenc} 
\usepackage[russian]{babel}   
\def\bc{\begin{center}}                \def\ec{\end{center}}
\def\beq{\begin{equation}}              \def\eeq{\end{equation}}
\def\bear{\begin{eqnarray}}            \def\eear{\end{eqnarray}}
\def\bt{\begin{tabular}}               \def\et{\end{tabular}}
\def\dst{\displaystyle}                \def\fns{\footnotesize}
\def\la{\langle}     \def\ra{\rangle}   \def\dg{\dagger}
       \def\lb{\label}     
\def\hs{\hspace}      \def\vs{\vspace}   \def\pr{\prime}
\def\sm{\small}       \def\pd{\partial}
  \def\rar{\rightarrow}
\def\td{\tilde}       \def\pr{\prime}
\def\noi{\noindent}   \def\nn{\nonumber}
\def\lrar{\longrightarrow}

\def\al{\alpha}      \def\bet{\beta}        
\def\Dl{\Delta}      \def\dl{\delta}    
\def\lam{\lambda}        \def\sg{\sigma}    
\def\om{\omega}          \def\t{\theta}

\begin{document}

\title{ {\sm \hfill E-print quant-ph/0408096 }\\[-3mm]
     {\sm \hfill Annual Higher Ped. Inst. Shumen, {\bf XIB}, 49-62
   (1987) \footnote{ Summary in Math. Review 1990k: 58078.}}\\[5mm]
 {\bf Geometric Quantization, Coherent States and
Stochastic Measurements}}
\author{B.A. Nikolov,\,  D.A. Trifonov } \date{} \maketitle

\begin{minipage}{130mm}{\sm
{\bf Abstract.} The geometric quantization problem is considered from the
point of view of the Davies and Lewis approach to quantum mechanics. The
influence of the measuring device is accounted in the classical and
quantum case and it is shown that the conditions of the measurement define
the type of quantization (Weyl, normal, antinormal, etc.). The quantum
states and quantum operators are obtained by means of the projection,
defined from the system of generalized coherent states. }

\end{minipage}
\vs{5mm}

\section{Introduction}

The main aim of the theory of geometric quantization consists in
construction of quantum-mechanical Hilbert space starting from the
properties of symplectic manifolds, which play the role of phase spaces of
classical mechanical systems [1,2]. If $G$ is a Lie group, then these
manifolds arise as orbits of co-adjoint action of $G$ in the dual space
$A(G)^*$ of the algebra $A(G)$ of the group $G$. Let $x_0\in A(G)^*$ be a
fixed element and $X$ be the orbit of $x_0$, $X=\{ax_0|\, a\in G\}$. At
some restrictions on $X$ [1,2], which we consider fulfilled (and therefore
classical system with phase space $X$ admits quantization), there exists a
unitary one-dimensional representation (a character) of a subgroup $K\subset
G$,
\beq\lb{1}
\chi(k) \equiv \chi(\exp(X_k)) = \exp\left(i\la x_0,X_k\ra\right),
\eeq
where $k \equiv \exp(X_k) \in K$ and $\la x_0,X_k\ra$ is the value of $x_0$
at the point $X_k\in A(K)\subset A(G)$. Then one can construct induced
representation  $W(G) = \chi(K)\!\uparrow\! G$, realized in the space
$H=L^2(X,dx)$ of square-integrable functions on $X$, where $dx$ is the
invariant measure on $X$, defined by means of the equality
\beq\lb{2}
\int_X dx\,f(x) = \int_G f(ax_0)da\, ,
\eeq
where $da$ is the invariant measure on $G$ (the group $G$ is supposed
unimodular).

Following Mensky [3], we consider the space $H$ as a space of virtual
states of quantum system, corresponding to classical system with phase
space $X$. The space of physical states appears as a subspace $H_0
\subset H$, in which the unitary irreducible representation $W_0(a) =
P_0W(a)P_0$ is realized, $P_0$ being projector on $H_0$.

In the present work we consider the relations between induced
representations and CS (Section 2) and show that the Hilbert space $H$ can
be treated as a space of classical states as well (Section 3). Then the
projector $P_0$ is interpreted as a quantization of classical system.
Next, in Section 4, we take into account the influence of the measuring
device, which allows different quantization rules to be treated in a
unified way. Finally, in Section 5,  we study the correspondence between
Poisson and Lie-algebraic structures.
\vs{3mm}

\section{Measurements, Coherent States and Induced \\
         Representations}

An unitary representation induced from the character $\chi(k)$ (1), acts in the space
$H$ according to the formula
\beq\lb{3}
(W(a)\Psi)(x) = \chi\left((a,x)_K\right)\Psi(a^{-1}x),
\eeq
where, by definition,
\beq\lb{4}
(a,x)_K = s^{-1}(x)a s(a^{-1}x).
\eeq
Here $s:\, G/K\rar G$ $-\!-$ cross-section of the Group fibre bundle
$(G,G/K,\pi_{\mbox{\tiny$G$}})$. We suppose that the space $X$ can be identified with
$G/K$, putting $aK = a x_0$, and thus $s:\, X\rar G$. The cross-section
$s$ is determined through decomposition of arbitrary element $a\in G$:
\beq\lb{5}
a = s(x_a)k_a = s(\pi_{\mbox{\tiny$G$}}(a))/k_a,\quad k_a\in K.
\eeq

Let $\om$ be some $K$-invariant vector in the space $H$, i.e.
\beq\lb{6}
W(k)\om = \lam(k)\om,\quad |\lam(k)| = 1.
\eeq
Herefrom, using (3) (at $a=k$ and $\Psi = \om$) we get
\beq\lb{7}
\chi((k,x)_K)\om(k^{-1}x) = \lam(k)\om(x).
\eeq

We construct system of states $\om_x = W(s(x))\om$ in $H$. The
representation $W$ transforms this system according to the formula
\bear\lb{8}
W(a)\om_x &=& W(as(x))\,\om  \nn \\
          &=& W\left(s(\pi_{\mbox{\tiny$G$}}(as(x))\right) k_{as(x)})\,\om
		   =   \lam\left((a,ax)_K\right)\om_{ax}\, .
\eear

Let us consider functions of the type $P_0\Psi$:
$$ (P_0\Psi)(x) = \la\om_x|\Psi\ra = \int dy\,
\overline{\om_x(y)}\,\Psi(y).\,  \eqno{(8a)} $$
Let $H_0  = \{\Psi\in H|\,\Psi=P_0\Psi\}$. Then if $\om\in H$ is such that
$$ W(k)\,\om = \chi(k)\,\om\, , \eqno{(6a)} $$
it is not difficult to verify that the subspace $H_0 \subset H$ is an
invariant subspace:
\beq\lb{9}
\bt{lll}
$\left(P_0W(a)\Psi\right)(x)
       $ &$=$& $\la\om_x|W(a)\Psi\ra = \la W(a^{-1})\om_x|\Psi\ra$\\[2mm]
$      $ &$=$& $\la\chi((a^{-1},a^{-1}x)_K)\,\om_{a^{-1}x}|\Psi\ra $\\[2mm]
$      $ &$=$& $\chi((a^{-1},a^{-1}x)_K^{-1}\Psi(a^{-1}x) $\\[2mm]
$      $ &$=$& $\chi((a,x)_K)\Psi(a^{-1}x) = \left(W(a)\Psi\right)(x)$\, .
\et \eeq
Now, using (3) and (6a), we obtain, that vector
$\om(x)$ satisfies the following condition: \footnote{one should not
mistake the function $\om(x)$ of the $K$-invariant vector $\om$ with
$\om_x:= W(s(x))\om$. (Note added). (There are no footnotes in the
journal paper $-\!-$ all are added in this electronic file in order to
clarify the exposition. No changes in the main text).} 
\beq\lb{10}
\om(k^{-1}x) = \chi((k,x)_K^{-1}k)\om(x)\, . 
\eeq
The states $\om_x$ form an overcomplete family of states in $H_0$, called
system of generalized coherent states (CS) [4]. Indeed, from (9) it
follows that
$ P_0 \om_x = \om_x,$
therefore
$$\la\om_z|\om_x\ra = \om_x(z) = \overline{\om_z(x)}. \eqno{(10a)} $$
Thus for every $\Psi \in H_0$ we have
\bear\lb{11}
\Psi(x) &=& (P_0\Psi)(x) = \la\om_x|\Psi\ra \nn \\
        &=& \int dz\,\overline{\om_x(z)}\Psi(z) = \int dz\,\om_z(x)\Psi(z)\,.
\eear
One can easily verify that $P_0$ is an orthoprojector:
\beq\lb{12}
\bt{ll}
$\dst (P_0^2\Psi)(x)$ & $\dst = \int dy\,\overline{\om_x(y)}\int\,dz
\overline{\om_z(x)}\Psi(z)$\\[2mm]
  & $\dst = \int dz \Psi(z)\la\om_x|\om_z\ra = \int dz\,\Psi(z)\om_z(x)
    = (P_0\Psi)(x),$\\[2mm]
$\dst  \la\Psi_1|P_0|\Psi_2\ra$ &$\dst = \la P_0\Psi_1|\Psi_2\ra\, .$
\et
\eeq

Let us turn now to the relationship between CS and covariant semi-spectral
measures (CSS-measures) [5].\footnote{Semi-spectral measures are also
called positive operator valued measures (POV-measures) [7,8]. Similarly,
spectral measure is a synonym of projection-valued (PV) measure.
(Note added).}
 Recall that the map $M:\, {\cal B}(X)\rar B(H)^+$
(where ${\cal B}(X)$ is a $\sg$-algebra of Borel subsets of $X$,\, $B(H)$
is algebra of bounded operators in $H$) is called CSS-measure associate
with the unitary irreducible representation $W(a)$, if the following
{\it covariance condition}  holds
\beq\lb{13}
W(a)M(\Dl)W^{-1}(a) = M(a\cdot\Dl),
\eeq
where $a\cdot\Dl = \{a\cdot x|\, x\in \Dl\}$, $\Dl \in {\cal B}(X)$.

If we denote the projector onto CS $\om_x$ as $M_x$,
\beq\lb{14}
(M_x\Psi)(y) = \om_x(y)\varphi(x),\quad \Psi \in H_0,
\eeq
then the set of operators
\beq\lb{15}
M_\om(\Dl) \equiv \int_\Dl dx\,M_x
\eeq
form a CSS-measure (in $H_0$).

The inverse turned out to be also true [6]: If $V(a)$ is an irreducible
square-integrable representation (of $G$) in some Hilbert space $H_0$, and
$M(\Dl)$ is a CSS-measure, then a finite trace operator $\rho_0$ exists, 
such that it commutes with all $V(k)$, $k\in K$, \footnote{ $K$ is
stationary subgroup of $x_0$ (see theorem IV.2.2 of [9], where CSS-measure
$\{M(\Dl)\}$ is shortly called covariant measurement. The invariant
measure $dx$ does not depend on the choice of $x_0$ [9]. (Note added).} and
\beq\lb{16}
M(\Dl) = \int_{\Dl} dx\,V(a)\rho_0\,V(a)^{-1},\quad (x = ax_0).
\eeq
In particular, if $M(\Dl)$ is an extremal point in the convex set of all
CSS-measures on $X$ and if ${\rm Tr}\rho_0 = 1$, then
\beq\lb{17}
\rho_0 =|\om_0\ra\la\om_0|,\quad\om_0 \in H_0,\,\, V(k)\om_0 = \chi(k)\om_0\,.
\eeq
In this case from (16) we again get (15), with $M_x =
V(a)|\om_0\ra\la\om_0| V(a)^\dg$. As we have seen in the above, condition
(17) means that $V(a)$ can be regarded as a subrepresentation of the
induced representation $W(a)$ ($W(a) = \chi(K)\uparrow G$) in the space 
$H = L^2(X)$.
The inclusion of $H_0$ in $H$ is determined by means of  the map $\Psi
\longmapsto \la\om_x|\Psi\ra$.  

If $W(G)$ is an induced representation, then CSS-measure $M(\Dl)$ appears
in a natural way from the canonical  spectral measure [6],
\beq\lb{18}
(\Pi(\Dl)\Psi)(x) = 1_\Dl(x)\Psi(x),
\eeq
where $$ 1_\Dl(x) = \left\{\bt{l} $1,\,\, x\in\Dl$\\ $0,\,\,
x\not\in\Dl$\et\right. , $$
for which the covariance condition (13) is valid too, and, in addition,
\beq\lb{19}
\Pi(\Dl_1)\Pi(\Dl_2) = \Pi(\Dl\cap \Dl_2)\, .
\eeq
(The last condition is not valid for an arbitrary CSS-measure). Then,
given a projector $P_0$ onto $H_0$, the operators
\beq\lb{20}
M(\Dl) = P_0\Pi(\Dl)P_0
\eeq
form a CSS-measure in $H_0$.

In the Davies and Lewis theory [7,8] the CSS-measure $M(\Dl)$ is generated
from the covariant instrument
\beq\lb{21}
E_\Dl(\rho) = \int_\Dl dx {\rm Tr}(M_x\rho)M_x
\eeq
using the relation
\beq\lb{22}
{\rm Tr} E_\Dl(\rho) = {\rm Tr}(\rho M(\Dl)).
\eeq
Let us note that the correspondence $\rho\lrar w_\rho$, where $w_\rho(\Dl) =
{\rm Tr}(\rho M(\Dl))$ is a measurement in the sense of Holevo
[6,9].\footnote{In [9] the set of operators $\{M(\Dl)\}$ is called
(generalized) resolution of unity. Since $\{M(\Dl)\}$ is in a one-to-one
correspondence with measurements (theorem II.2.1 of [9]) they are shortly
called measurements. (Note added).}

If $f(x)$ is a real continuous function on the phase space $X$, then one
can define operators
\beq\lb{23}
M(f) = \int dx f(x) M_x,
\eeq
and
\beq\lb{24}
E_f(\rho) = \int dx f(x) {\rm Tr}(M_x\rho)M_x\, ,
\eeq
from which $M(\Dl)$ and $E_\Dl$ are obtained as particular cases at $f(x)
= 1_\Dl(x)$. When $f(x)$ is interpreted as classical observable, then
$M(f)$ could be regarded as a quantum observable, corresponding to the
classical one. When $M_x$ is a projection on a Glauber CS the so defined
correspondence $f\lrar M(f)$ is called {\it stochastic quantization} [10].
Operator $E_f(\rho)$, Eq. (24), is interpreted as a (unnormalized) state,
to which the system goes after measurement of the observable $M(f)$ [7,8].


\section{Algebras of Classical Observables and their Quantization}

Let $C_0(X)$ be algebra of finite continuous functions on $X$ (with usual
multiplication and complex conjugation as an involution). Then, using (3)
and introducing the representation $\Pi:\, C_0(X)\lrar B(H)$,
\beq\lb{25}
(\Pi(f)\Psi)(z) = f(z)\Psi(z),
\eeq
one can easily check the covariance condition
\beq\lb{26}
W(a)\Pi(f)W(a)^\dg = \Pi(L_af),
\eeq
where $(L_af)(x) = f(a^{-1}x).$ The relation between $\Pi(f)$ and spectral
measure $\Pi(\Dl)$ is given by the expressions
\beq\lb{27}
\Pi(1_\Dl) = \Pi(\Dl),
\eeq
\beq\lb{28}
\Pi(f) = \int f(x)\Pi(dx).
\eeq
If one denotes the density of the measure $\Pi(\Dl)$ as $\Pi_x$, then (28)
could be rewritten in the form
$$\Pi(f) = \int dx\, f(x)\Pi_x\, , \eqno{(28a)} $$
where \footnote{ $\dl_x(z)$ is the Dirac $\dl$-function. (Note added).}
\beq\lb{29}
(\Pi_x\Psi)(z) = \dl_x(z)\Psi)z)\, .
\eeq
Note that the operators $M_x$ in (14) and $\Pi_x$ in (29) are related as
follows
\beq\lb{30}
P_0\Pi_xP_0 = M_xP_0\, ,
\eeq
wherefrom
\beq\lb{31}
P_0\Pi(f)P_0 = M(f)P_0\, .
\eeq
In this way, treating the elements of $C_0(X)$ as classical observables,
the representation of algebra $C_0(X)$ in the space $H = L^2(X)$ is
determined by means of Eq. (28) in complete analogy with Eq. (23). Then in
correspondence with (30), (31) the transition from classical to quantum
observables  is represented by the projector $P_0$, Eq. (8a), which maps
$H$ onto subspace $H_0$:
\beq\lb{32}
(P_0\Psi)(x) = \int dy\,\la\om_x|\om_y\ra\Psi(y)\, .
\eeq
Now if the states of classical system are characterized by means of
density matrices $\rho$ in $H$, i.e. $\rho(x,y) = \overline{\rho(y,x)}$,
Tr$\rho:= \int dx\,\rho(x,x) = 1$, then the mean value of observable $f$,
Eq. (28a), in a state $\rho$ would be
\beq\lb{33}
\la f\ra = {\rm Tr}\left(\rho\Pi(f)\right) = \int dx\,f(x)\rho(x,x)\, .
\eeq
In particular, for $\rho(x,y) = \Psi(x)\overline{\Psi(y)} =
(\Psi\otimes\bar{\Psi})(x,y)$ we can write (33) in form of the matrix
element $\la\Psi|\Pi(f)\Psi\ra$,
\beq\lb{34}
\la f \ra_\Psi = \equiv \la f \ra_{\Psi\otimes\bar{\Psi}} =
\int dx\overline{\Psi(x)}f(x)\Psi(x) = \la\Psi|\Pi(f)\Psi\ra\,.
\eeq
Apparently $\rho(x,x)$ is a probability distribution density describing
the preparation of classical system.

Let us note that the description of classical instrument by means of
a formula like (24),
\beq\lb{35}
E^{(cl)}_f(\rho) = \int dx\,f(x){\rm Tr}\left(\rho\Pi_x\right)\Pi_x,
\eeq
is possible iff the space admits a reproducing kernel $K(x,y)$,
for which
\beq\lb{36}
\Psi(x) = \int dy\, K(x,y)\Psi(y),\quad    K(x,x)=1\,.
\eeq
Then
$$ (\Pi_x\Psi)(y) = \dl_x(y)\Psi(y) = \int dz\,\dl_x(y)K(y,z)\Psi(z), $$
and therefore ${\rm Tr}\Pi_x = \int dz\,\dl_x(z)K(z,z) = K(x,x) =1$.

\section{Observables Conditioned by Measuring Device}

Following Prugovecki [11] we accept that an exhausting description of the
measurement result "$x$" is attainable if one introduce a nonnegative function
$\eta_x$ with a maximum at $x$, interpreted as a probability
distribution density of true values $z$ of the measuring quantity in space
$X$. Then the quantity
\beq\lb{37}
\eta_{\mbox{\tiny$\Dl$}}(z) = \int_\Dl dx\,\eta_x(z),\quad \eta_{\mbox{\tiny $X$}}(z) = 1,
\eeq
determines the probability of obtaining the true value $z$, if the
measurement yielded value $x$ in Borel set $\Dl$, $\Dl \in{\cal B}(X)$.
In this way the function $\eta_{\mbox{\tiny$\Dl$}}$ describes statistical
error introduced by the measuring device. In accordance with the Holevo's
classical statistical model conjectures [9] we suppose that the probability
distribution of a given measurement in a state $w(dx) = w(x)dx =
\rho_w(x,x)dx$  is given by the formula
\beq\lb{38}
\mu_w(\Dl) = \int w(dx)\,\eta_{\mbox{\tiny$\Dl$}}\,.
\eeq
Then the mean value of observable $f(x)$ in the measurement (in Holevo
sense) $w\,\longmapsto \mu_w$ is equal to
\beq\lb{39}
\la f\ra_{\mu_w} \equiv \int f(x)\mu_w(dx) = \int w(dx)f_\eta(x) = \la
f_\eta\ra_w,
\eeq
where
\beq\lb{40}
f_\eta(x) = \int dz\,f(z)\eta_z(x)
\eeq
would be called {\it classical observable, conditioned by $\eta$-device},
or $\eta$-observable. Then, in view of $f_{\eta=\dl}(x) = f(x)$, function
$f(x)$ could  be called $\dl$-observable. Let us note that
$\dl$-observables could be rather arbitrary if the device-functions
$\eta_z(x)$ are well behaved. If for example, $\eta_z\in C_0(X)$ for all
$z\in X$ and $f\in C(X)$, then $f_\eta \in C_0(X)$. We denote the set of
$\dl$-observables as $a(X)$. Supposing $\eta_z \in C_0(X)$ we put
$$\Pi_\eta(f) = \Pi(f_\eta) = \int dz\, f(z)\Pi_{\eta_z} =
\int f(z)\Pi_\eta(dz)\, ,$$
where the operators  $\Pi_\eta(\Dl)$, $\Pi_{\eta_z}$ are generalizations of
 $\Pi(\Dl)$, $\Pi_z$, Eqs. (18), (29), defined by means of relations
\beq\lb{41}
\bt{l}
$\dst (\Pi_{\eta}(\Dl)\Psi)(x) = \eta_{\mbox{\tiny$\Dl$}}(x)\Psi(x),$\\[2mm]
$\dst (\Pi_{\eta_z}\Psi)(x) = \eta_z(x)\Psi(x)$ .
\et \eeq

Let us suppose that the device-functions $\eta_z$ obey the relation
\beq\lb{42}
\eta_{az}(x) = \eta_z(a^{-1}x).
\eeq
Then one can straightforwardly show that $\Pi_\eta(\Dl)$ is a CSS-measure:
\beq\lb{43}
W(a)\Pi_\eta(\Dl)W(a)^{-1} = \Pi_\eta(a\cdot\Dl).
\eeq
Herefrom it also follows that
\beq\lb{44}
W(a)\Pi_\eta(f)W(a)^{-1} = \Pi_\eta(L_a f)\, ,
\eeq
where (similarly to Eq. (25))
$$(\Pi_\eta(f)\Psi)(z) = f_\eta(z)\Psi(z). $$

The transition to quantum observables is performed in a manner similar to
that described in Section 3, and is expressed in terms of the map
\beq\lb{45}
\bt{l}
$\dst \Pi_{\eta_z} \lrar P_0\Pi_{\eta_z}P_0 \equiv M_{\eta_z}$,\\[2mm]
$\dst M_{\eta_z} = \int dx\,\eta_z(x)M_x$\, .
\et
\eeq
Then quantum observable that corresponds to the classical one $f(z)$ will
be (the operator)
\beq\lb{46}
M_\eta(f) = \int dz\,f(z)M_{\eta_z} = M(f_\eta)\, .
\eeq

The quantization rule (46) generalizes the Davies--Lewis stochastic
quantization (23) and allows one to consider the known quantization rules
[12] in a unified manner.

Consider for example the case of the usual (two dimensional) phase space
$R^2(q,p) \approx C(z)$, $z=(q+ip)/\sqrt{2\hbar}$. In view of (see [12])
\footnote{In this example  $d^2z/\pi$ is the invariant measure. (Note added).}
$$M_z = |z\ra\la z| = \int \frac{d^2\al}{\pi}
\exp\left(\al\bar{z}-\bar{\al}z + \frac{\al\bar{\al}}{2}\right)D(\al),$$
$$D(\al) = \exp(\al \hat{a}^\dg -\bar{\al}\hat{a}),\quad
[\hat{a},\hat{a}^\dg] = 1\, ,$$
we have
$$M_\eta(f) = \int (d^2z/\pi)\,f(z)\int
(d^2\al/\pi)\td{\eta}_z(\al)\,e^{-\al\bar{\al}/2} D(\al),$$
where
$$\td{\eta}_z(\al) = \int
(d^2z/\pi)\,\eta_z(\al)\exp(\al\bar{z}-\bar{\al}z)\, .$$
Now choosing $\eta_z(\al)$ such that $\td{\eta}_z(\al) =
\exp(\al\bar{\al}/2-\al\bar{z}-\bar{\al}z)$, we obtain the well known Weyl
quantization rule:
$$ M_w(f) = \int (d^2z/\pi)f(z)\hat{W}(z) = \int (d^2\al/\pi)f(\al)D(\al),
$$
where $\hat{W}(z) = \int (d^2\al/\pi)\exp(\al\bar{z}-\bar{\al}z)D(\al)$ is
the Wigner operator, the mean of which in a (mixed) state $\hat{\rho}$ is
equal to the Wigner function [13].

Let us turn now toward the algebraic structure of the set of classical and
of quantum $\eta$-observables.

The product $\ast$ of the quantum observables we define by means of the
relation
\beq\lb{47}
{\rm Tr}\left(\rho \left(M_\eta(f)\ast M_\eta(g)\right)\right) =
{\rm Tr}\left(E_{f_\eta}(E_{g_\eta}(\rho))\right)\, ,
\eeq
which is a generalization of an analogical relation, proposed by F.E.
Schroeck [14] in his dequantization program in the framework of Devies and
Lewis theory. Using (47) we may put 
\beq\lb{48}
M_\eta(f)\ast M_\eta(g) = \int dx\int dz\,f_\eta(x)g_\eta(z)\bet(x,z)M_z\, ,
\eeq
where $\bet(x,z) = |\la\om_x|\om_z\ra|^2$. If we define a new product
$\ast$ of classical observables by means of the relation
\beq\lb{49}
(f\ast g)_\eta(z) = \int dx\,f_\eta(x)g_\eta(z)\bet(x,z)\, ,
\eeq
we could treat $M_\eta$ as a homomorphism of classical observable algebra
$a(X)$ into the algebra of quantum observables:
\beq\lb{50}
M_\eta(f\ast g) = M_\eta(f)\ast M_\eta(g)\, .
\eeq

The classical limit is understood as [14]:
$$ \bet(x,z) \lrar \dl_z(x). $$
Then we have
$$ (f\ast g)_\eta(z)|_{\bet\rar \dl} = f_\eta(z)g_\eta(z) =
(f \ast_c\, g)_\eta(z),$$
where the product  $\ast_c $, defined by means of the relation
$(f \ast_c\,g)_\eta(z) = f_\eta(z)g_\eta(z)$ allows as to treat
$\Pi_\eta$ as a homomorphism in the algebra of classical observables,

\beq\lb{51}
\Pi_\eta(f \ast_c\,g) = \Pi_\eta(f)\Pi_\eta(g)\, .
\eeq

In this way both classical and quantum observables can be regarded as
functions on $X$ with different multiplication rules, $\ast_c$
and $\ast$, the multiplication $\ast$  coinciding with
$\ast_c$ in the limit $\bet(x,z) \rar \dl_z(x)$.\footnote{\, Unlike
$\ast_c$ the multiplication $\ast$ is non-associative and non-commutative.
Let us note that $\Pi(fg)=\Pi(f)\Pi(g)$, which is to be compared with (51).
(Note added).}

\section{Lie Algebraic Structure of Algebras of Observables}

It is well known that the dynamics of classical systems can be described
by means of canonical transformations. Let $X$ be a space with symplectic
structure $\om = \om_{ij}dx^i\wedge dx^j$, and let its first cohomology
group $H^1(X)$ be trivial (that is every closed 1-form $\t = \t_idx^i$,
$d\t=0$, on $X$ is exact: $\t = df$). Then the infinitesimal canonical
transformation of space $X$ has the form \footnote{\, $\tau$ is an external
parameter (the time). Here a transformation is called canonical if it
preserves the symplectic structure. (Note added).}
\beq\lb{52}
x^i \lrar x^{\pr\,i} = x^i + \dl\tau\om^{ij}\pd g/\pd x^j\, ,
\eeq
where $\om^{ij}$  is a matrix inverse  to $\om_{ij}$, and $g$ is a smooth
function, $g\in C^\infty(X)$. The transformation (52) induces a
transformation of classical observables, which are assumed to be also in
the class of $C^\infty$:
\bear\lb{53}
 f\lrar f^\pr:\,\, f^\pr(x) &=& f(x^\pr) = f(x+\dl\tau\om\pd g/\pd x) \nn \\
 &=& f(x) + \dl\tau\om^{ij}\frac{\pd g}{\pd x^i}\frac{\pd f}{\pd x^j} =
 f(x) + \dl\tau\{f,g\} \nn \\
 &=& f(x) - \dl\tau X(g)f\simeq (e^{-\dl\tau X(g)}f)(x), 
 \eear
 where $\{f,g\}$ is the Poisson bracket,
$$\{f,g\} = \om^{ij}\frac{\pd f}{\pd x^i}\frac{\pd g}{\pd x^j}, \eqno{(53a)}$$
and $X(g)$ is the generator of the canonical transformation. The
transformation (53) can be represented in the differential form
\beq\lb{54}
\frac{\pd f}{\pd\tau} + X(g)f = 0\, ,\quad\mbox{\small where}\quad
X(g) = \om^{ij}\frac{\pd g}{\pd x^i}\frac{\pd}{\pd x^j}\, .
\eeq

On the other hand, using the phase volume invariance under canonical
transformations (Liouville theorem) and the relation
\beq\lb{55}
\frac{\pd}{\pd x^i}\left(\om^{ij}\frac{\pd g}{\pd x^j}\right) = 0
\eeq
one can obtain the identity [15]
$$ \int dx\,\left(e^{-\dl\tau X(g)}f\right)(x)w(x) =
   \int dx\,f(x)\left(e^{-\dl\tau X(g)}w\right)(x)\, .$$
Herefrom it follows that under canonical transformations the observables
could be regarded as invariant, provided the states (understood as
probability distributions $w(x)$\,)  vary according to the law \footnote{\,
Eq. (56) is the Liouville equation in general coordinates $x^i$,
corresponding to Hamilton function $g(x)$ (and $X(g)$ is the Hamilton
vector field). (Note added).} 
\beq\lb{56}
\frac{\pd w}{\pd \tau} = X(g)w\, .
\eeq
Putting $w(x) = |\Psi(x)|^2$, where $\Psi\in H$, one easily verify that
equation (56) is a consequence of the equation
\beq\lb{57}
\frac{\pd \Psi}{\pd \tau} = X(g)\Psi\, ,
\eeq
which could be regarded as "classical Schr\"odinger equation" with
Hamiltonian $iX(g)$.

By direct calculation we obtain the transformation law of the
operator $\Pi(f)$ (Eq. (25)) which represents in $H$ the algebra of classical
observables:
\beq\lb{58}
e^{\dl\tau X(g)}\Pi(f)e^{-\dl\tau X(g)} =  \Pi\left(e^{\dl\tau
X(g)}f\right)\, .
\eeq
Herefrom
\beq\lb{59}
[X(g),\Pi(f)] = \Pi(X(g)f)\, ,
\eeq
and, in view of
\beq\lb{60}
[X(g),X(f)] = X(\{g,f\})\, ,
\eeq
the pair $(\Pi,X)$ can be regarded as a representation of Poisson algebra
$\la C^\infty_0(X),\, \cdot \, , \, \{\,,\,\}\ra$ [16] in Hilbert space $H$.

In order to pass to the algebra of $\eta$-observables we introduce a new
Poisson bracket $\{\,*\!\!_, \,\}$  such that
\beq\lb{61}  %
\{g *\!\!\!_, \, f\}_\eta = \{g_\eta,f_\eta\}\, ,
\eeq
where we suppose that $\eta_x\in C^\infty_0(X)$ for all $x\in X$. Then if
one put
\beq\lb{62}
X_\eta(g) = X(g_\eta)\, ,
\eeq
one can get relations, similar to (59), (60),
\beq\lb{63}
[X_\eta(g),X_\eta(f)] = X_\eta(\{g *\!\!\!_,\, f\})\, ,
\eeq
\beq\lb{64}
[X_\eta(g),\Pi_\eta(f)] = \Pi_\eta(\{g *\!\!\!_,\, f\})\, .
\eeq
The relations (51), (63), (64) show that the pair $(\Pi_\eta,X_\eta)$ form
a representation of the Poisson algebra of $\eta$-observables
$\la a_\eta(X),\, \ast_c \, , \, \{\,*\!\!_,\,\} \ra$ in Hilbert space $H$.

The quantization is performed by means of projection onto subspace $H_0$.
Introducing the notations $Q(f)$ and $Q_\eta(f)$,
\beq\lb{65}
Q(f) = P_0X(f)P_0,\quad Q_\eta(f) = Q(f_\eta),
\eeq
we  have
\beq\lb{66}
[Q_\eta(f),Q_\eta(g)] = P_0X_\eta(f)P_0 X_\eta(g)P_0 -
P_0X_\eta(g)P_0 X_\eta(f)P_0.
\eeq
Suppose now that the commutation relation
\beq\lb{67}
X(\eta_x)P_0 = P_0 X(\eta_x)
\eeq
is valid for all $x\in X$. Then from (66) and (63) it follows that
\beq\lb{68}
[Q_\eta(f),Q_\eta(g)] = Q_\eta(\{f *\!\!\!_,\, g\}\,.
\eeq
Similarly, from (64) we have
$$[Q_\eta(f),M_\eta(g)] = M_\eta(\{f *\!\!\!_,\, g\}\,. \eqno{(68a)}$$
Recalling now the relation (50) we see that the pair $(M_\eta,Q_\eta)$
can be regarded as a homomorphism of Poisson algebra
$\la a(X),\, \ast_c \, , \, \{\,*\!\!_,\,\} \ra$ in the algebra of
operators with multiplication $*$, Eq. (48). One has to note that the
relation between the Poisson bracket and the commutator (68) does not
require any limit transition.

The relation (67), which in coordinates has the form
\beq\lb{69}
\om^{ij}(x)\frac{\pd \eta_{x^\pr}(x)}{\pd x^i}\frac{\pd\la\om_x|\om_y\ra}
{\pd x^j} + \om^{ij}(y)\frac{\pd \eta_{x^\pr}(y)}{\pd y^i}
\frac{\pd\la\om_x|\om_y\ra} {\pd y^j} = 0\, ,
\eeq
establishes a connection of the device functions $\eta_{x^\pr}$ to the
reproducing kernel $\la\om_x|\om_y\ra$ in the space $H_0$ (in derivation
of (69)  the condition (55) has been also used).
\vs{3mm}

{\Large Acknowledgment.} One of the authors (D.T.) is grateful to S.T.
Ali and F.E. Schroeck for stimulating and fruitful discussions.

\vs{15mm}

{\Large\bf References}
\begin{itemize}
\def\im{\item}

\im[{[1]}] Kostant, B.\,  $-\!-$ \, Lect. Notes Math., Vol. 170 (1970).

\im[{[2]}] Souriau, J.M.\, $-\!-$ \, {\it Structure des syst\'emes dynamique},
Dunod, Paris, 1970.

\im[{[3]}] Mensky, M.B.\, $-\!-$ \, Commun. Math. Phys., Vol. 47, p. 97 (1976).

\im[{[4]}] Perelomov A.M.\, $-\!-$\, Commun. Math. Phys., Vol. 26, p. 222 (1972).

\im[{[5]}] Scutaru, H.\, $-\!-$ \, Lett. Math. Phys., Vol. 2, p. 101 (1977).

\im[{[6]}] Holevo, A.S.\, $-\!-$\, Repts. Math. Phys., Vol. 16, p. 385
(1979).

\im[{[7]}] Davies E.B., J.T. Lewis\, $-\!-$\, Commun. Math. Phys., Vol. 17,
      p. 239 (1970).

\im[{[8]}] Ali S.T., E.G. Emch \, $-\!-$\, J. Math. Phys., Vol. 15, p. 176 (1974).

\im[{[9]}] Holevo A.S.\, $-\!-$\, {\it Probabilistic and Statistical Aspects of
       Quantum Theory} (in Russian). "Nauka", Moscow, 1980. \footnote{\, 
	   English translation: North-Holland, Amsterdam, 1982.}

\im[{[10]}] Schroeck, F.E.\, $-\!-$\, Found. Phys., Vol. 12, p. 825 (1982).

\im[{[11]}] Prugovecki, E.\, $-\!-$\, J. Math. Phys., Vol. 17, p. 825 (1982).

\im[{12]}] Agarwal, G.S., E. Wolf\, $-\!-$ Phys. Rev. D, Vol. 2, p. 2161 (1970).

\im[{[13]}] Wigner, E.\, $-\!-$\, Phys, Rev., Vol. 40, p. 749 (1932).

\im[{[14]}] Schroeck, F.E.\, $-\!-$\, {\it The Dequantization Programme for
Stochastic Quantum Mechanics}, Florida Atlantic Univ. Preprint,
1983. \footnote{\, J. Math. Phys. {\bf 26}, 306-310 (1985). (Note added).} 

\im[{[15]}] Balescu, R.\, $-\!-$\, {\it Equilibrium and Nonequilibrium
Statistical Mechanics}, Wiley, New York, 1975.

\im[{[16]}] Lichnerovicz, A.J.\, $-\!-$\, Diff. Geometry, Vol. 12, p. 253
(1977); Guillemin, V., S. Sternberg,\, $-\!-$\, Ann. Phys., Vol. 127, p. 220
(1980).

\end{itemize}
\newpage
.
\vs{-23mm}
\bc
 \framebox(155,230){ 
\begin{minipage}{13cm}{\sm 
\bc
ГОДИШНИК НА ВИСШИЯ ПЕДАГОГИЧЕСКИ ИНСТИТУТ 

В ШУМЕН 
\vs{-3mm}
\ec
\centerline{Том XI, Б \hs{25mm} Природо-математически факултет \hs{25mm} 1987} 

\bc ANNUAL OF THE HIGHER PEDAGOGICAL INSTITUTE 

    IN SHUMEN 
\vs{-3mm}
\ec
\centerline{Volume XI, B \hs{28mm} Faculty of Sciences \hs{46mm} 1987} 	
\vs{3mm}

\hrule
\vs{10mm}

\bc {\sm ГЕОМЕТРИЧЕСКОЕ КВАНТОВАНИЕ, КОГЕРЕНТНЫЕ}\\
    {\sm СОСТОЯНИЯ И СТОХАСТИЧЕСКИЕ ИЗМЕРЕНИЯ}
\ec
\vs{3mm}
\bc {\sl Благовест Николов,\,\,  Димитър Трифонов} \ec
\vs{5mm}
 
 1. ВВЕДЕНИЕ
\vs{2mm}
 
 Основная цель теории геометрического квантования состоит в
конструировании квантовомеханического пространства Гильберта, исходя из
свойств симплектических многообразий, играющих роль фазовых пространств
классических механических систем [1,2]. Если $G$ - группа Ли, то эти
многообразия возникают как орбиты коприсоединенного действия $G$ в
дуальном пространстве $A(G)^*$ алгебры $A(G)$ группы $G$. Пусть $x_0 \in
A(G)^*$ -- фиксированный элемент и $X$ -- орбита точки $x_0$, $X=\{ax_0| 
a\in G\}$. При некоторых ограничениях на многообразие $X$ [1,2], которые
мы считаем выполненными (и значит классическая система с фазовым
пространством $X$ допускает квантование), существует унитарное одномерное
представление (характер) подгруппы $K\subset G$,
\vs{4mm}

$\dst(1)$\hs{30mm} $\dst\chi(k)\equiv \chi(\exp(X_k))=\exp(i\la x_0,X_k\ra)$,
\vs{4mm}

\noi
где $k\equiv \exp(X_k) \in K$ и $\la x_0,X_k\ra$ значение $x_0\in A(G)^*$
в точке $X_k\in A(K)\subset A(G)$. Тогда можно построить индуцированное
представление $W(G) = \chi(K)\uparrow G$, реализующееся в пространстве $H
= L^2(X,dx)$ квадратично интегрируемых функций на $X$, где $dx$
инвариантная мера на $X$ .... 
\vs{2mm}

 .......................
\vs{2mm}
   
В настоящей работе мы рассматриваем связь между индуцированными
представлениями и когерентными состояниями (п. 2) и показываем, что
пространство Гильберта $H$ можно рассматривать также в качестве
пространства классических состояний (п. 3). Тогда проектор $P_0$
интерпретируется как квантование классической системы. Далее (п. 4) мы
учитываем влияние измерительного прибора, что позволяет рассматривать
известные правила квантования с единных позиций. Наконец (п. 5) мы изучаем
соответствие между пуассоновскими и Ли-алгебрическими структурами. } 
\vs{10mm}

$^4$ {\fns Годишник на ВПИ-Шумен, Т. XI, Б}
\end{minipage}
}
\ec
 
\end{document}